\documentclass[aps,prb,twocolumn,superscriptaddress]{revtex4}
\usepackage{graphicx}
\usepackage{epsfig}
\usepackage{float}

\newcommand{\be}{\begin{equation}}
\newcommand{\ee}{\end{equation}}
\begin{document}

\title{Sr$_2$Cu(PO$_4$)$_2$: A real material realization of the 1D nearest neighbor Heisenberg chain}
\author{M.D. Johannes}
\affiliation{Code 6390 Naval Research Laboratory, Washington, D.C. 20375}
\author{J. Richter}
\affiliation{Institute for Theoretical Physics, University of Magdeburg, Germany}
\author{S.-L. Drechsler}
\affiliation{Leibniz-Institute for Solid State and Materials Research (IFW Dresden), Dresden, Germany}
\author{H. Rosner}
\affiliation{Max-Planck-Institute for Chemical Physics of Solids, Dresden, Germany}

\begin{abstract} We present evidence that crystalline Sr$_2$Cu(PO$_4$)$_2$ is a nearly perfect one-dimensional (1D) spin-1/2
anti-ferromagnetic Heisenberg model (AHM) chain compound with nearest neighbor only exchange.  We undertake a broad theoretical study
of the magnetic properties of this compound using first principles (LDA, LDA+U calculations), exact diagonalization and Bethe-ansatz
methodologies to decompose the individual magnetic contributions, quantify their effect, and fit to experimental data.  We calculate
that the conditions of one-dimensionality and short-ranged magnetic interactions are sufficiently fulfilled that Bethe's analytical
solution should be applicable, opening up the possibility to explore effects beyond the infinite chain limit of the AHM Hamiltonian.  
We begin such an exploration by examining some extrinsic effects such as impurities and defects.  \end{abstract}

\maketitle

\section{Introduction} Interest in magnetically low-dimensional systems began with the advent of quantum mechanics and the
development of spin-spin interaction models to explain magnetic behavior. Though deceptively simple, early models such as
Ising \cite{ising} or Heisenberg \cite{heis} cannot be solved for the general case, requiring either low spin or low
spatial dimensionality to obtain analytical solutions.  Exact solutions for some specific cases, such as Onsager's solution
\cite{onsager} to the two-dimensional (2D) Ising model or Bethe's solution \cite{Bethe} to the spatially one-dimensional
(1D) s=1/2 Heisenberg chain, inspired general interest in the theoretical properties of magnetism in low dimensions, and
resulted in a variety of predictions for exotic physical behaviors.  Some of the most interesting properties to arise from
the study of restricted-dimension magnets are due to the dramatic effect of quantum spin fluctuations \cite{qm}.  These are
intimately involved in the emergence of novel ground states and excitation spectra, and, in recent decades, have become a
leading candidate as the pairing mechanism for electrons in quasi-2D high temperature superconductors \cite{JPC99}, thus
providing the field with a more practical aspect.  Mermin and Wagner demonstrated \cite{Mermin-Wagner} that strong spin
fluctuations suppress magnetic long range order (LRO) until T=0 in isotropic spin systems in 1D and 2D, 
regardless of the interaction
strength (exchange)  between neighboring spins.  Since real physical compounds are 3D by nature, the continuing effort to
experimentally verify such predictions and probe the nature of 1D magnets is aimed at finding materials where magnetic
interactions proceed predominantly along one direction.  A measureable gauge of success or failure along
these lines can be obtained through the ratio $k_BT_N/J_1$, which compares ordering temperature of a N\'eel state to the
magnitude of exchange between neighboring spins.  A perfectly 1D system would give $k_BT_N/J_1$ = 0.  Additionally,
experimentally observed behaviors should conform to theoretical predictions where the conditions of the model are met.

Recently, very low temperature measurements \cite{beliknew} on the spin-1/2 compound Sr$_2$Cu(PO$_4$)$_2$
identified the onset of magnetic LRO at T$_N$= 0.085 K, which, in combination with the extracted exchange
constant, yields a ratio $k_BT_N/J_1$ = 6x10$^{-4}$.  This is nearly a full order of magnitude less than the
ratio \cite{moto} for prototype 1D magnet Sr$_2$CuO$_3$, which has $k_BT_N/J_1$ = 2.5x10$^{-3}$.  We can
estimate the remaining interchain coupling, $J'$, by adopting an effective 3D chain model with $z_{\perp}$=4
nearest neighbor chains as in Ref. \onlinecite{irkhin00}: 

\begin{equation} J'=\frac{3.046k_BT_N}
{k_{AFM}z_{\perp}\sqrt{\ln \frac{5.8J_1}{k_BT_N} +0.5\ln \ln\frac{5.8J_1}{k_BT_N}}} \approx 2.9\times
10^{-2}\mbox{K}, \label{qcc} \end{equation} 

where $k_{AFM}$ is the magnitude of the AFM wavevector.  All signs, therefore point to an extremely high degree of
one-dimensionality that should justify the use of Bethe's exact analytical solution to the 1D spin-1/2 AHM in a
wide temperature range $J' \ll k_BT$, provided that indeed {\it only nearest-neighbor} interactions are present.  
Previous studies \cite{helge} have shown that this condition is not satisfactorily fulfilled by the leading 1D
spin-1/2 AHM candidate, Sr$_2$CuO$_3$.  For completeness, another candidate for a 1D-AHM system should be
mentioned:  the linear charge transfer salt [3,3'-dimethyl-2,2'-thiazolinocyanine]-TCNQ
\cite{takagi96,remarktakagi}.

Here, we undertake an extensive theoretical analysis of Sr$_2$Cu(PO$_4$)$_2$, employing first principles density
functional theory calculations, an exact diagonalization scheme, and finally, a Bethe-Ansatz derived expression
for the magnetic susceptibility fit to experimental data.  Our results are in remarkable agreement with one
another and with experimental observations.  We show that Sr$_2$Cu(PO$_4$)$_2$ is indeed extremely 1D and that
furthermore, the second-neighbor interactions are vanishingly small, eliminating any complications due to in-chain
frustration.  We claim, therefore, that this compound is the best realization of a nearest-neighbor only
Heisenberg spin 1/2 chain known to date and will likely be a valuable tool for extracting intrinsic effects beyond
the Bethe-ansatz {\it i.e.} effects not contained in the Heisenberg Hamiltonian, such as Dzyaloshinsky-Moriya
interactions \cite{dzy}, ring exchange processes, or spin-lattice coupling effects.  Additionally, extrinisic
effects due to sample imperfections such as defects, magnetic impurities or the presence of alternate phases can
be quantitatively explored, and we make a preliminary investigation of some of them.

\begin{figure}[tbp] \begin{center} \includegraphics[width = 0.95\linewidth]{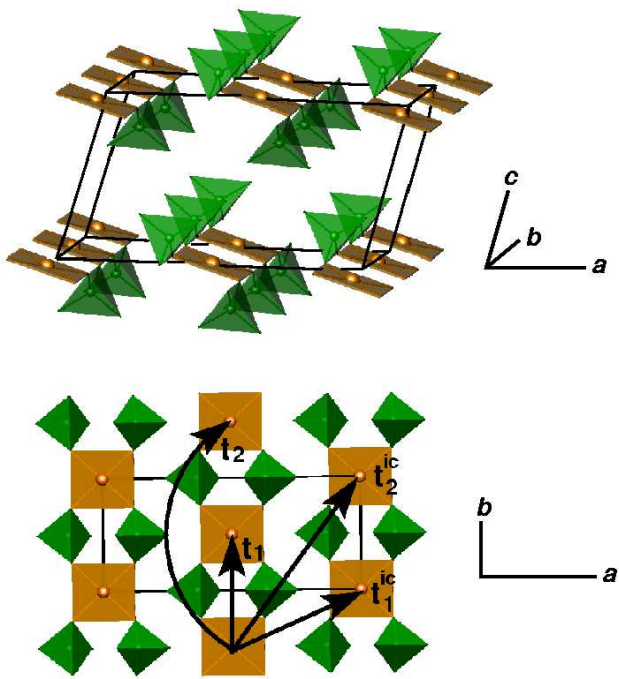} 
\end{center} \caption{(Color online) {\it 
Upper panel} The structure of Sr$_2$Cu(PO$_4$)$_2$, showing the isolated CuO$_4$ plaquettes as 
squares with O ions located at the corners, and the $PO_4$ units as tetrahedra with $P$ located 
in the center. {\it Lower panel} A top view of the spin chain plane with the various hopping 
paths labelled.  Hopping to the nearest out-of-plane neighbor, t$_{\perp}$, is not shown.} 
\label{struct} \end{figure}

\begin{figure}[b]
\begin{center}
\includegraphics[width=0.95\linewidth]{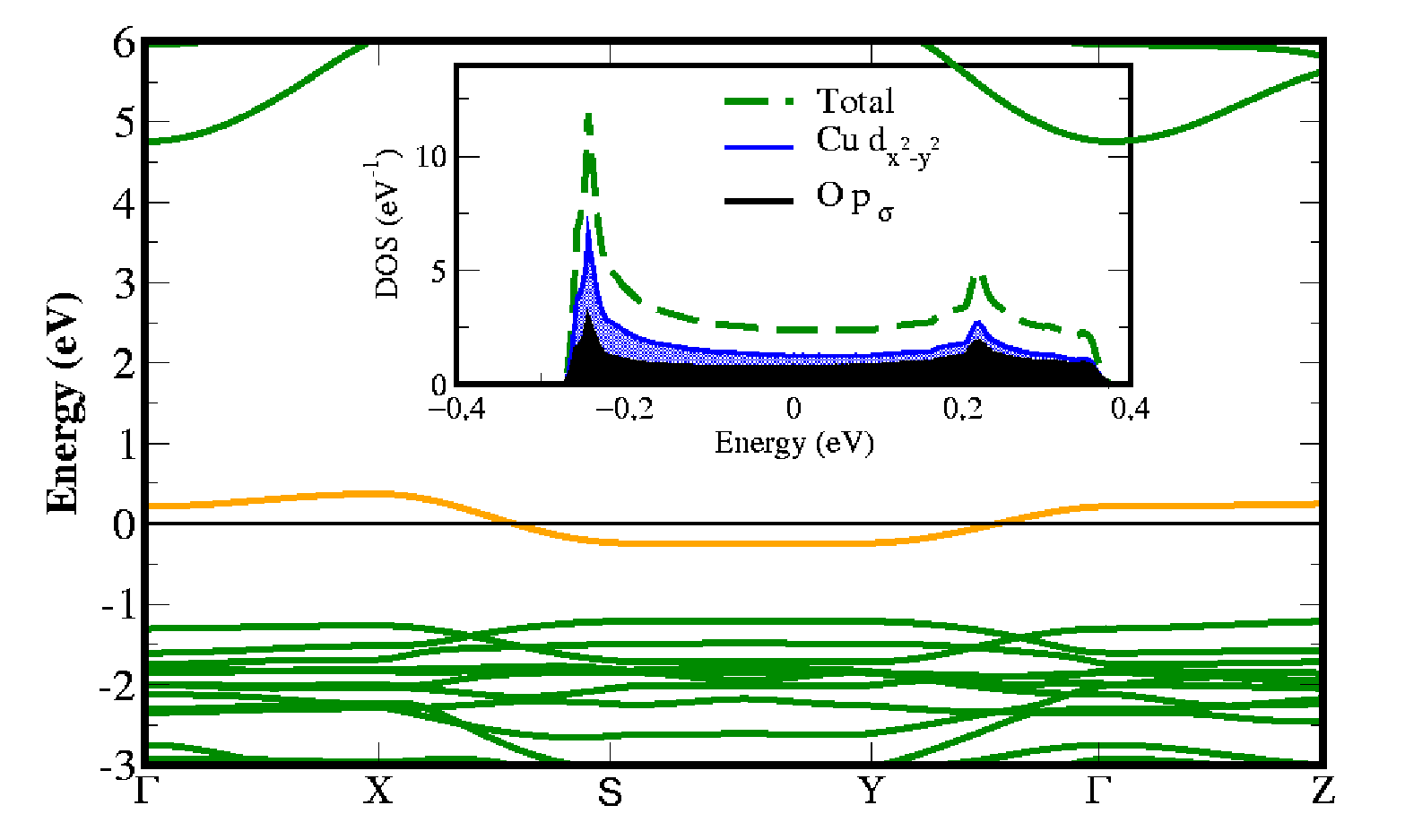}
\includegraphics[width=0.95\linewidth]{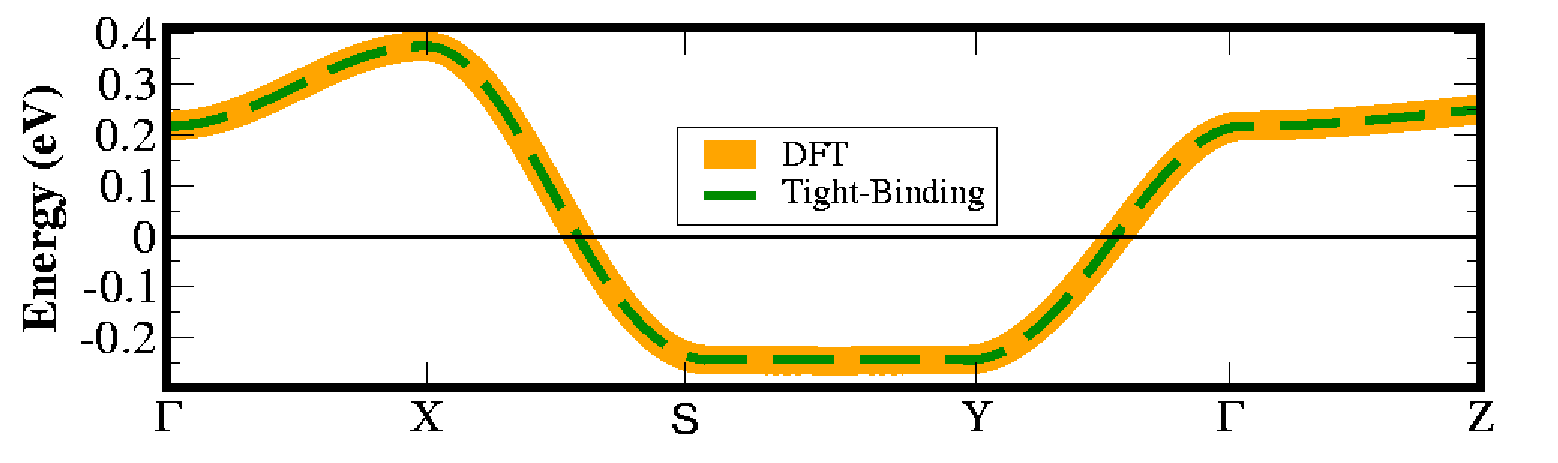}
\end{center}
\caption{(color online) The bandstructure of Sr$_2$Cu(PO$_4$)$_2$, showing the single metallic band
well separated from all others.  The $X-S$ and $Y-\Gamma$ directions are along the magnetic chain, while 
the $S-Y$ and $\Gamma-Z$ directions are 
perpendicular to the chain. The inset shows the total and orbitally resolved DOS for the single
band. In the lower panel, a 
blow-up of this band is shown with tight-binding eigenvalues superimposed to show 
the remarkable reproduction of the dispersion.  
} \label{BS} \end{figure}

\section{First Principles and Tight-Binding} First principles density functional theory (DFT) bandstructure
calculations were performed using a full-potential local orbital code, FPLO \cite{fplo}, with the following
valence states:  Sr (4s,4p,4d,5s,5p), Cu (3s,3p,3d,4s,4p), P (2s,2p,3s,3p,3d), O (2s,2p,3d).  The structure,
lattice constants, and atomic positions (see Fig. \ref{struct}) were taken from experiment
\cite{belik1,AAB+02}, $a$ = 11.515 \AA, $b$ = 5.075 \AA, $c$ = 6.5748 \AA. 
Sr$_2$Cu(PO$_4$)$_2$ and isostructural Ba$_2$Cu(PO$_4$)$_2$, which we calculate as a
comparison material, are composed of isolated CuO$_4$ plaquettes surrounded by PO$_4$ tetrahedra.  The
plaquettes are spaced evenly along the $b$ axis, forming chains that are staggered with respect to one
another in the $a$ crystal direction.  The planar CuO$_4$ units are tilted with respect to the $a-b$ crystal
plane.  Based on the geometry of these two systems, we enumerate five specific interactions between
plaquettes, and therefore between spins localized to these plaquettes, that may be necessary to describe the
electronic and magnetic structures.  These interactions are labelled in Fig. \ref{struct} as various hopping
parameters to be later included in a tight-binding model.

The paramagnetic band structure of Sr$_2$Cu(PO$_4$)$_2$ (Fig. \ref{BS}) shows a single, isolated, half-filled band,
derived predominantly from the Cu $3d_{x^2-y^2}$-O $2p_{\sigma}$ molecular plaquette orbital, crossing the Fermi
energy.  In reality, the system is antiferromagnetic (AFM) and insulating, but we will follow the standard procedure
of importing hopping parameters from the "uncorrelated" paramagnetic system to the Hubbard model which then maps onto
the Heisenberg model.  Magnetism and correlation effects can also be added at the DFT level with, as we will show,
very consistent results.  The 1D character of the system is qualitatively obvious from the nearly dispersionless
bands in directions perpendicular to the magnetic chains (S-Y and $\Gamma$-Z) and from the characteristic logarithmic
divergences in the density of states (DOS) near the band edges.  To quantitatively compare microscopic magnetic
interactions, we fit a tight-binding (TB) model to our paramagnetic band structure and calculated the individual
exchange constants between various CuO$_4$ plaquette spins using J$_{ij}$ = 4t$_{ij}^2$/U$_{eff}$ with U$_{eff}$=4.5
eV.  We find this to be a reasonable choice for U$_{eff}$ because of poor screening in this geometry and because of
the small inter-plaquette repulsion. The hopping parameters included in the model are shown
schematically in Fig. \ref{struct} with the numerical values and derived superexchange constants given in Table
\ref{tbparam}. The resulting TB dispersion, which uses only 5 fitting parameters, is indistinguishable from the
full-potential calculation (Fig. \ref{BS} lower panel), indicating that further interactions can be safely ignored.  
The ratio of the strongest in-chain coupling to the strongest interchain coupling is J$_1$/J$_1^{ic}$ $\sim$ 70 and
the ratio of first to second neighbor in-chain coupling is J$_1$/J$_2$ $\sim$ 700 for Sr$_2$Cu(PO$_4$)$_2$. Identical
calculations based on the band structure of Ba$_2$Cu(PO$_4$)$_2$ (not shown) yield similar results with somewhat more
inter-chain coupling but less second neighbor in-chain coupling.  Both systems can therefore be considered as
strongly one-dimensional, with Ba$_2$Cu(PO$_4$)$_2$ slightly less so.  Naturally, the choice of U$_{eff}$ is simply a
best estimation and results for $J$ will vary slightly with this choice, while the ratios will remain constant.

\begin{table}[h] \caption{Tight-binding hopping parameters (in units of meV) and derived exchange constants (in units 
of K) for A$_2$Cu(PO$_4$)$_2$, A=Sr,Ba.  See Fig. 
\ref{struct} for the
relationship of the hopping parameters to the structure.}
\begin{tabular*}{0.95 \linewidth}{@{\extracolsep{\fill}}||ccccccccccccc||} \hline \hline
&(meV) & \vline & t$_1$ & \vline & t$_2$ & \vline & t$_1^{ic}$ & \vline & t$_2^{ic}$ & \vline & t$_{\perp}$ & \\ \hline
& Sr & \vline & 135 & \vline & 5.1 & \vline & 16.3 & \vline & 3.4 & \vline & 1.4 &\\ \hline \hline
& Ba & \vline & 122 & \vline & 0.9 & \vline & 10.3 & \vline & 4.7 & \vline & 1.8 & \\ \hline \hline
\end{tabular*}

\vspace{0.05 in}

\begin{tabular*}{0.95 \linewidth}{@{\extracolsep{\fill}}||ccccccccccccc||} \hline \hline
 &(K)&  \vline & J$_1$ & \vline  & J$_2$ & \vline & J$_1^{ic}$ & \vline & J$_2^{ic}$ &\vline & J$_{\perp}$ &    \\ \hline
& Sr & \vline &  187 & \vline &  0.268 & \vline & 2.7 & \vline & 0.119  & \vline & 0.02 &\\ \hline \hline
& Ba & \vline & 154 & \vline &  0.008 & \vline & 1.09 & \vline & 0.228 & \vline & 0.03 & \\ \hline
\hline               \end{tabular*}

\label{tbparam}
\end{table}

The energy difference between FM and AFM ordered spin configurations can be calculated using the local spin density approximation 
(LSDA) \cite{JPP92} which allows for separate spin-up and spin-down densities.  Since the LSDA is known to badly underestimate the 
onsite Coulomb interaction in localized systems, we applied the LSDA+U methodology to better account for the correlated Cu 
$3d$-orbitals, using the fully localized limit scheme \cite{via93} to correct for double-counting terms.  We map a classical N\'eel 
state and a classical ferromagnetic state onto the Heisenberg spin model, including only 1D nearest neighbor interactions.  
Comparing the resulting model energy difference to the LDA energy difference between FM and AFM states (per spin), we derive an 
effective exchange constant, J$_{eff}$, in the following way:

\begin{equation}
H = \sum_{i,j} J_{ij} \mathbf{S}_i \cdot \mathbf{S}_j; \quad  E_{FM} - E_{AFM} =
2J_{eff}|s|^2 ,\quad s=1/2,
\label{heis}
\end{equation}

As expected, the energy difference, and therefore $J_{eff}$, decreases as U$_d$ (not to be confused with the 
considerably smaller one-band 
parameter U$_{eff}$ that contains O 2$p$ contributions in addition to Cu $3d$)  increasingly localizes the Cu $3d$ electrons. For a 
range of U$_d$ between 6 eV and 9 eV, we find that $J_{eff}$ varies from 261 K to 160 K.  Since U$_d$ is a local 
quantity 
and since the Cu-O 
bond distance in Sr$_2$Cu(PO$_4$)$_2$ is only 1\% different than in the plaquettes of the widely studied high T$_c$ precursor systems, 
we adopt the commonly used value of U$_d$ = \mbox{8 eV}.  This corresponds to a value of 190 K for $J_{eff}$. Note that 
since the LSDA (and LSDA+U) 
energy differences include contributions from {\it all} exchange processes in the system, $J_{eff}$ cannot in general be considered as 
either purely superexchange or purely 1D.  However, comparison with the individual superexchange parameters derived from the TB fit 
shows that both assumptions, in this case, are quite valid. The nearest neighbor in-chain TB exchange constant has a value $J_1$ = 187 
K, in exceptional agreement with the $J_{eff}$ value of 190 K, indicating that the next-nearest neighbor interactions, FM exchange 
processes and residual 2D and 3D interactions must therefore be extremely small. Of course both values can be made to vary somewhat by 
choosing U$_d$ and U$_{eff}$ differently, thus affecting the agreement as well.  We expect that our calculated value 
of $J$ will be larger 
than the experimental value, as it is well known that the band dispersion from which we derive $t$ and subsequently $J$ is generally 
exaggerated by the LDA.  Indeed, the experimentally derived value of $J$ is 143 K, in good but not perfect agreement with our 
calculations.  We emphasize that any renormalization of the hopping parameters stemming from effects outside the LDA will cancel in the 
ratio ($J_1$/$J_2$) so that the precise calculated value of $J$ has, in any case, no bearing on the establishment of the compound's 
pronounced magnetic one-dimensionality and short-ranged magnetic interactions that are the primary aim of our first principles study.

\section{Exact Diagonalization} 
\begin{figure}
\includegraphics[width=0.95\linewidth]{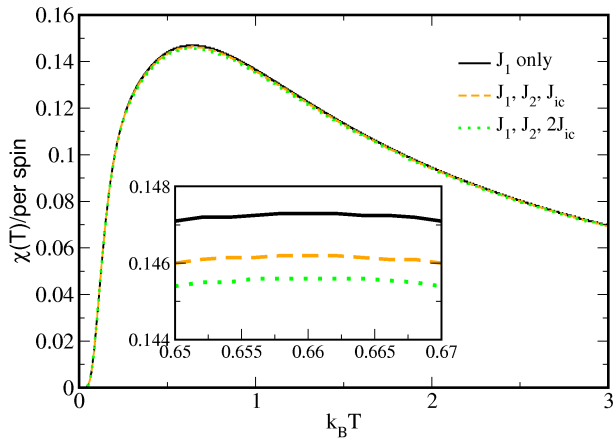}
\caption{A 20 site exact diagonalization calculation of the susceptibility per spin of Sr$_2$Cu(PO$_4$)$_2$.  The 
inset shows a blow-up of the
region where the curves have the greatest discrepancy - 0.9\%.  Curves for the ground state correlation of spins and for the specific
heat per spin show even less deviation.}
\label{ed}
\end{figure}

We perform an exact diagonalization calculation using ten sites along two staggered AFM magnetic chains 
(20 sites total) for calculating thermodynamic properties such as specific heat and magnetic susceptibility, and 
using 36 total sites for obtaining the ground state properties.  
We compare calculations including the three largest exchange interactions, $J_1$, $J_2$ and $J_{ic}^1$, as 
listed in Table \ref{tbparam}, 
to calculations using only $J_1$.  Since in our model each chain has only one neighboring chain, we also perform a calculation in which 
$J_{ic}^1$ is doubled to account for the existence, in reality, of two neighboring chains.  In the case of specific heat, there is no 
discernable difference between any of the three curves using these different parameter sets.  For the calculation of $\chi$, the curves 
are identical for the majority of the temperature range explored ($ 0 < k_bT/J_1 < 3$), but a barely visible difference occurs near the 
peak of the curve (see Fig. \ref{ed}).  The maximal difference occurring between two curves is 0.9\%, at about 
$k_BT/J_1$=0.65.  The 
ground state calculations of the spin-spin correlation, $\langle S_i \cdot S_j\rangle$ are again completely indistinguishable.  
Obviously, more distant and weaker interactions, such as $J_{ic}^2$ and $J_{\perp}$ will have even less of an effect.  From this we 
conclude that Sr$_2$Cu(PO$_4$)$_2$ is essentially free of magnetic interactions beyond the first neighbor and is an ideal candidate for 
Bethe-ansatz calculations, which we now discuss.

\section{Comparison with experiment}
The basic theory for non-periodic, open AHM
chains was worked out by Furusaki and Hikihara \cite{Furu} (FH) and also by Zvyagin and Makarova \cite{Zvy} (ZM). FH 
considered a
half-infinite chain with $one$ free chain end applying bosonization theory whereas ZM considered finite even-membered chains on the
basis of a rigorous theory based on the Bethe ansatz. Both approaches result, at low temperature, in a chain length ($L=Na$) 
dependent
 diverging contribution to the total magnetic susceptibility
$\chi \propto 1/NT \ln(T_0J_1$)
and to the linear coefficient in the specific heat $\gamma
=C_p/T \propto 1/NT \ln^\beta (T_0J_1/T)$, with $\beta$=2,4 in the FH and ZM theory,
respectively. Since a real chain has two ends we multiplied the FH expressions
by a factor of 2. ZM calculated further logarithmic corrections which
we adopted here to be valid
for the FH case, too.
Then within both approaches we arrive finally at the same expression for the
chain end contributions to
$\chi$ (up to a factor of 3/4). For the logarithmic constant $T_0$
we used the same value 5.696 as proposed by Johnston {\it et al.} in
the fit expression (fit2) for the bulk susceptibility \cite{johnston},
similarly to 5.8 used in Eq.\ (1).
In the shown and described fits we have adopted the formalism of FH, modified as described above, for both $C/T$ and $\chi$.

\begin{figure}[tbp]
\includegraphics[width= 0.95 \linewidth]{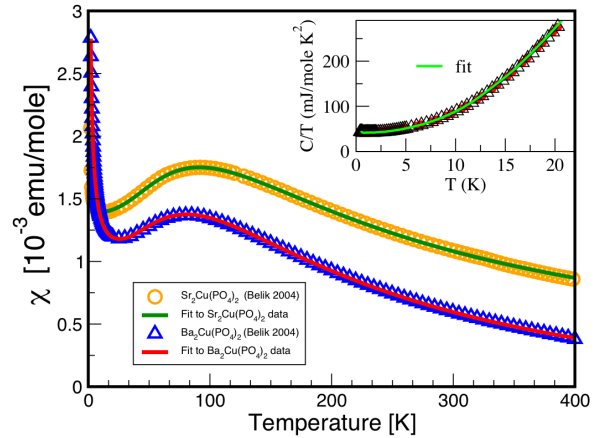}
\caption{Bethe-Ansatz fits to experimental data with corrections for broken chains, impurity phases, and domain boundaries.  Note the 
consistency of the exchange parameter vs. the wide spread in values for the chain length.}
\label{BA}
\end{figure}

\begin{table}
\caption{The collected results of fits to different sets of $\chi (T)$ data for Sr$_2$Cu(PO$_4$)$_4$ and Ba$_2$Cu(PO$_4$)$_4$.}
\begin{tabular*}{0.95 \linewidth}{@{\extracolsep{\fill}}|c|c|c|c|} \hline \hline
Sr$_2$Cu(PO$_4$)$_2$ &J(K) & N & g \\ \hline \hline
Belik \cite{belik1} & 144.9  &114 &2.713 \\ \hline
Belik \cite{beliknew} & 143.9 & 337.7 & 2.154 \\ \hline
Nath \cite{loidl} & 152.6 & 5 & 1.952  \\ \hline
\end{tabular*}

\vspace{0.05 in}

\begin{tabular*}{0.95 \linewidth}{@{\extracolsep{\fill}}|c|c|c|c|} \hline \hline
Ba$_2$Cu(PO$_4$)$_2$ &J(K) & N & g \\ \hline \hline
Belik \cite{belik1} & 131.8  & 48 & 2.073 \\ \hline
Nath \cite{loidl} & 140.1 & 5 & 2.041  \\ \hline
\end{tabular*}
\end{table}

We fit to both susceptibility and specific heat to data taken on the same sample \cite{belik1}, additionally using data from a later 
sample \cite{loidl} for which only $\chi$ data was available.  The fit quality for specific heat and susceptibility are very similar; 
results for the latter are shown in Fig.  \ref{BA}.  We get very good fits throughout the entire temperature range and find a reasonably 
consistent value for the exchange parameter, $J$, despite fitting to samples of different qualities and to two different measurements 
($C/T$ and $\chi$).  It is worth noting that the exchange parameter derived from fitting to $C/T$ using the data of Ref. 
\onlinecite{belik1} yields an exchange parameter of J=134.4 K, which is somewhat less than the value derived from fitting to $\chi (T)$ 
using the {\it same} sample.  This is likely caused by non-magnetic impurity contributions (not accounted for in our model) that affect 
the specific heat but not the susceptibility at low temperatures.  Since the magnetic component of $C$ is $\propto 1/J$, ignoring 
the 
non-magnetic contribution overestimates this term and therefore underestimates the exchange.  Unlike the exchange and $g$ values, the 
chain length parameter, $N$, varies widely between samples. While this is expected for samples of different quality, the variation is 
surprisinly high and, more importantly, the chain lengths resulting from our fits are far too small to justify the use of the open chain 
theories that we have employed at very low temperatures.  With chain lengths of this order, the low temperature region will be 
completely dominated by broken chain physics that requires different, and as-of-yet undeveloped, formalism.  Having fit throughout a 
large temperature range, including regions where broken chain physics is inoperative, we feel that the extracted exchange constant, J 
$\approx$ 145 K, is nonetheless relevant - a belief that is supported by its consistency between fits and its similarity to the 
experimentally measured value.

\section{Discussion} The application of a variety of theoretical techniques to the problem of magnetism in Sr$_2$Cu(PO$_4$)$_2$
convincingly demonstrates that the ideal compound is highly one-dimensional. The high degree of one-dimensionality can be traced
back to its unusual isolated CuO$_4$ plaquette geometry. Instead of edge- or corner- shared plaquettes such as are common in
other quasi-1D compounds \cite{KMK+97,YMTT+98}, each Cu ion in Sr$_2$Cu(PO$_4$)$_2$ is surrounded by four O ions not shared by
any other Cu ion.  This construction virtually eliminates the second neighbor in-chain coupling that prevents edge-shared
compounds such as Li$_2$CuO$_2$ from being described via a simple nearest neighbor Heisenberg model \cite{Warren,Rosner}.  
Corner shared cuprates such as Sr$_2$CuO$_3$ have far smaller second neighbor interactions \cite{moto}, of the order J$_1$/J$_2$
$\sim$ 15, and yet, these must be taken into account to get good agreement between model calculations and experiment
\cite{helge}.  The structure of Sr$_2$Cu(PO$_4$)$_2$ along the chain is that of an edge shared chain compound with every other
unit missing. Conceptualized in this way, one can make a correspondence between exchange constants in a edge-shared ($es$) system
and those in the isolated square plaquette ($sp$) geometry: J$_2^{es}$ $\rightarrow$ J$_1^{sp}$, and J$_4^{es}$ $\rightarrow$
J$_2^{sp}$.  Since J$_4^{es}$ is known to be vanishingly small in the edge-shared geometry, it is clear that the second neighbor
interactions in the square plaquette geometry can be expected to be negligible. This may provide some directional guidance in the
search for new one-dimensional compounds: the isolated plaquette arrangement appears to be superior to the more common edge- or
corner-shared structures such that synthesis of new compounds with this geometry may prove to be profitable. The tilting of the
out-of-chain plaquettes with respect to one another further suppresses the inter-chain coupling.  The staggering of plaquettes in
neighboring chains slightly increases the distance between spins, but more importantly, gives rise to frustration.  As each chain
is antiferromagnetically aligned by the (relatively) strong first neighbor coupling, a given spin finds itself surrounded by four
interchain neighbors, two aligned in one direction and two in the other.  These staggered, frustrated chains are more decoupled
from one another than they would be in another arrangement, {\it e.g.} a ladder configuration.

Provided that Sr$_2$Cu(PO$_4$)$_2$ is stoichiometric and largely defect-free, it is clear that this compound represents the most
1D AHM chain so far investigated.  These conditions are, unfortunately, not reasonably fulfilled by current samples.  It is
interesting in this context to consider the mechanism by which the compound eventually achieves LRO (at $T_N$ = 0.085 K): is it
truly the result of residual third dimension interactions?  Significantly, the phenomenologically estimated averaged interchain
interaction from Eq. \ref{qcc} is of the same order as the calculated $J_{\perp}$, rather than $J_{ic}^1$.  Fluctuation induced
"order by disorder" coupling could be responsible for the strong reduction of two orders of magnitude within the frustrated plane.  
On the other hand, the interchain couplings are in general phenomenally small as calculated by DFT methods and even so are likely
exaggerated.  One alternative explanation is that in a system with many broken chains, there will be some number of chains
containing an odd number of spins, with each such chain carrying one uncompensated spin-1/2 electron.  The relationship of the
uncompensated spins to one another is not defined by any of our methods and a long range ordering of these is not out of the
question.  It would be interesting to see if the ordering temperature remains constant with sample quality.  Another point to be
addressed in the future is the issue of spin-lattice coupling.  The Heisenberg model itself assumes perfect isotropy in spin-space
and we have not included any relativistic (spin-orbit) interactions in our first principles calculations.  The neglect of these is
seemingly justified by the extremely small field (H= 4mT) at which the spin-flop transition occurs \cite{beliknew}, but the
smallness of this field itself is unusual and a cause for further investigation. All of these facts point to the high desirability
of better samples that can be used to disentangle true "dimensionality" effects from behaviors due to crystal imperfections.
Although the investigation of true Heisenberg physics is currently limited by sample quality issues, Sr$_2$Cu(PO$_4$)$_2$ is 
theoretically, and potentially experimentally, the best example of a magnetically 1D crystal yet studied.   

\section{Conclusion} We have shown that the isolated CuO$_4$ plaquette geometry of Sr$_2$Cu(PO$_4$)$_2$ gives
rise to a nearly perfect 1D spin-1/2 nearest neighbor only system.  We find a ratio $k_BT_N/J_1$ = 6x10$^{-4}$,
in good agreement with experimental finding and show that secondary interactions (2D,3D and next-nearest
neighbor) are negligible in terms of calculated thermodynamic properties.  Using the Bethe-ansatz solution to the
Heisenberg Hamiltonian along with additional terms to correct for extrinsic non-crystalline effects, we fit the
data over a large temperature range and derive an exchange parameter of 145 $\pm$ 5 K that is consistent between
samples and between fitting choices.  We find that sample quality, particularly the existence of numerous broken
chains, currently prohibits experimental observation of true spin-1/2 AHM physics.  However, Sr$_2$Cu(PO$_4$)$_2$
is truly intrinsically perfectly 1D with only one exchange parameter, and as better and better methods of
generating the compound emerge, effects beyond Bethe-ansatz can be probed experimentally.

{\it Acknowledgements} We are grateful to A.A. Belik and A. Loidl for making data available to us and for
valuable input to our work.  We also thank A.A.  Zvyagin for fruitful discussions.  We acknowledge the use of
J.~Schulenburg's {\it spinpack} to perform the numerical exact diagonalization.  Our calculations were carried
out in part using the supercomputing facilities at ZIH Dresden. Funding was provided by the Emmy Noether Program
of Deutsche Forschungsgemeinschaft.  MDJ is funded by the Office of Naval Research.


\begin{thebibliography}{27}
\bibitem[{rem()}]{remarktakagi}
\bibinfo{note}{However, for this compound exhibiting the a steep decrease of
  $\chi(T)$ at low $T$ the ratio $T_N/J_1$ amounts about 2.3$^.$10$^{-2}$,
  only. In addition, the reported fit of the infinite chain Bethe-ansatz
  solution without any subtraction of Curie-contributions shows significant
  deviations near the maximum of $\chi (T)$ and at high $T$.}
\end{thebibliography}
\end{document}